\documentclass[a4paper]{article}

\usepackage[T1, T2A]{fontenc}
\usepackage[utf8]{inputenc}
\usepackage[english]{babel}

\usepackage[a4paper,top=3cm,bottom=2cm,left=3cm,right=3cm,marginparwidth=1.75cm]{geometry}

\usepackage{amsmath}
\usepackage{graphicx}
\usepackage[colorinlistoftodos]{todonotes}
\usepackage[colorlinks=true, allcolors=blue]{hyperref}
\usepackage{authblk}

\title{Discrete Shilnikov attractor and chaotic dynamics in the system of five identical globally coupled phase oscillators with biharmonic coupling}
\author[1]{\small Evgeny A. Grines}
\author[2,1]{\small Alexei O. Kazakov}
\author[3]{\small Igor R. Sataev}
\affil[1]{\footnotesize Lobachevsky   State   University   of   Nizhni   Novgorod,   23
Gagarin av., Nizhny Novgorod 603950, Russia}
\affil[2]{\footnotesize National Research University Higher School of Economics, 25/12 Bolshaya Pecherskaya Ulitsa, 603155 Nizhny Novgorod, Russia}
\affil[3]{\footnotesize Kotelnikov's Institute of Radio-Engineering and Electronics of RAS, Saratov Branch, Zelenaya 38, Saratov, 410019, Russia}
\affil[ ]{\textit{evgenij.grines@gmail.com}}
\date{}

\begin{document}
\maketitle

\begin{abstract}
{
We argue that a discrete Shilnikov attractor exists in the system of five identical globally coupled phase oscillators with biharmonic coupling. We explain the scenario that leads to birth of this kind of attractor and numerically illustrate the sequence of bifurcations that supports our statement.\\\\}
\end{abstract}


\section*{Introduction}

Systems of interacting elements attract a great interest of specialists in nonlinear dynamics and dynamical systems. The resulting collective behaviour of such systems could be quite complex \cite{PikRos2015,StankPerMcClintStefa2017} even in the case when a dynamics of a single element is very simple. One of the approaches for studying such systems is to approximate them by systems of coupled phase oscillators. For example, Winfree model \cite{Winfree1967} describes how a population of weekly coupled limit cycle oscillators can be described by a system of coupled phase oscillators.

Systems of identical phase oscillators are of special interest here. In systems of identical elements the coupling between oscillators becomes the main source of complex behaviour. For example, choosing coupling function that has only the first Fourier harmonic leads to Kuramoto or Kuramoto-Sakaguchi system. While they were widely used in studying and explaining synchronization, not all of collective behaviour phenomena are possible to observe in this system; for example, it is not possible to have multiple clusters \cite{EngelMir2014,PikRos2015}. However, adding a second harmonic\footnote{Note that there are also other ways to get chaotic attractors in systems of identical globally coupled phase oscillators. This could be done by using coupling function with higher order harmonics \cite{AshBickTimm2011} or by taking into account nonpairwise interactions between phases \cite{AshBickRodr2016}.} makes dynamics much richer, allowing formation of clusters \cite{Okuda1993}, heteroclinic cycles \cite{HansMatoMeunier1993,kori2001slow} and chaotic attractors\footnote{This is in contrast with a case of non-identical oscillators where chaotic attractors can emerge even in the simplest Kuramoto model due to strong detuning of oscillators' natural frequencies \cite{PopMaistrTass2005}.} \cite{AshTownOrosz2007}. Thus, we have examples of chaotic attractors in systems of identical globally coupled phase oscillators and it is very natural to ask: how do chaotic attractors in these systems emerge? What bifurcation scenarios lead to an appearance of strange attractors? 

The present work studies some of the routes that lead to an emergence of strange attractors in the system of five identical globally coupled phase oscillators with biharmonic coupling \cite{AshTownOrosz2007}. It is important to note that studying the five-dimensional flow that corresponds to this system can be reduced to studying some three-dimensional Poincar{\'e} map (see Section \ref{num_bif} for details). There are many papers devoted to studying chaotic dynamics in three-dimensional maps and we want to highlight, in particular, \cite{GonGonShil2012, GonGonKazTur2014, GonGon2016}. These papers present several phenomenological scenarios for an emergence of homoclinic attractors\footnote{An attractor is called homoclinic if it contains a saddle fixed point\cite{GonGon2016}.} in one-parameter families of three-dimensional maps. The first two scenarios lead to an emergence of either a discrete Lorenz attractor or a figure-eight attractor depending on the multipliers of a saddle fixed point. These attractors appear due to the following sequence of bifurcations. First, an attracting fixed point undergoes the period doubling bifurcation and becomes a saddle. Then the point of period 2 that emerged from previous bifurcation loses its stability and after that a homoclinic structure to the saddle appears. Note that these two scenarios have no analogues in continuous time dynamical systems.

The idea of the third scenario from \cite{GonGonShil2012, GonGonKazTur2014} dates back to the Shilnikov's paper \cite{Shilnikov86}. In this paper the scenario for an emergence of homoclinic attractor of spiral type is described. This attractor is based on a homoclinic trajectory to a saddle-focus equilibrium. The scenario has following stages. First, a stable equilibrium undergoes Andronov-Hopf bifurcation and becomes a saddle-focus. As a result of this a stable limit cycle appears. After that the unstable manifold of the saddle-focus starts to wind around this limit cycle and forms a "funnel"-type configuration. The limit cycle loses its stability later and after that a homoclinic trajectory to the saddle-focus appears. We say that {\it Shilnikov attractors} are the attractors that emerge according to this scenario.
Shilnikov attractors can be found in many different systems, including R{\"o}ssler system, Arneodo-Coullet-Tresser systems \cite{ArnCoulTres1980, ArnCoulTres1982}, and Rosenzweig-MacArthur system \cite{RosMcArth1963, KuznFeoRin2001, BakhKazKor2017}. 
The scenario that we described above was translated later to the case of three-dimensional mappings \cite{GonGonShil2012, GonGonKazTur2014} (see description in Section \ref{scenario}). 
Attractors of spiral type that appear in accordance with this scenario are called {\it discrete Shilnikov attractors}.
First examples of such attractors were found in three-dimensional generalized H{\'e}non maps \cite{GonGonShil2012, GonGonKazTur2014}. Regarding the systems from applications, the only previous example that is known to us is a homoclinic Shilnikov attractor in a Poincar{\'e} map for the nonholonomic model of Chaplygin top \cite{BorKazSat2016}.

In present paper we will show that a discrete Shilnikov attractor exists in the system of five identical globally coupled phase oscillators with the biharmonic coupling  \cite{AshTownOrosz2007}. This attractor emerges according to the scenario from \cite{GonGonShil2012}. The paper is organized as follows. In section \ref{scenario} we describe the scenario of a discrete Shilnikov attractor appearance. In section \ref{num_bif} we introduce the system from \cite{AshTownOrosz2007} and present the numerical analysis that verifies the presence of discrete Shilnikov attractor.

\section{Scenario of a discrete Shilnikov attractor emergence}
\label{scenario}

Let us describe the scenario for an emergence of a homoclinic attractor based on a saddle-focus fixed point in a three-dimensional map.
Consider a one-parametric family of three-dimensional maps $X_{n+1} = F(X_n, \mu)$ and suppose that the following sequence of bifurcations is observed. 
For $\mu < \mu_1$ the fixed point $O_{\mu}$ is an attractor. At $\mu = \mu_1$ the fixed point $O_{\mu}$ undergoes Neimark-Sacker bifurcation\footnote{This bifurcation is also sometimes called a torus birth bifurcation or a discrete Andronov-Hopf bifurcation.} 
and the stable invariant torus $L_{\mu}$ emerges. After this bifurcation the fixed point becomes a saddle-focus of type (1, 2), i.e. it has one-dimensional stable manifold $W^s(O_{\mu})$ and two dimensional unstable manifold $W^u(O_{\mu})$. At $\mu = \mu_2 > \mu_1$ the invariant curve $L_{\mu}$ changes its type from nodal to focal. As a result, the invariant manifold $W^u(O_{\mu})$ starts to wind around the invariant curve $L_{\mu}$ and thus creates a ''funnel'' that attracts all trajectories starting in the neighbourhood of the saddle-focus $O_{\mu}$ except the points at its stable manifold $W^u(O_{\mu})$. Suppose that if we further increase parameter $\mu$ then the stable manifold $W^s(O_{\mu})$ intersects the unstable manifold $W^u(O_{\mu})$. Without loss of generality we may assume that these intersections happen on the interval $\mu_2 < \mu_3 < \mu < \mu_4$. The presence of homoclinic trajectories alone implies complicated dynamics \cite{Smale65, Shil67, Newhouse79} and it doesn't depend on transversality of intersection. According to \cite{GonGonShil2012}, if the invariant curve $L_{\mu}$ is destroyed\footnote{Multiple scenarios of torus breakdown and emergence of strange attractor are known to date. Torus breakdown can happen due to Afraimovich-Shilnikov scenarios 
or because of a cascade of torus doublings. Note that it is still unknown whether this cascade is finite or infinite. It is often observed that chaotic behaviour appears after a finite number of torus doublings \cite{ArnColSp1983, Anish}.}
and there is no regular attractors at its place on the interval $(\mu_3, \mu_4)$ or earlier, then a discrete Shilnikov attractor emerges. This attractor is based on a homoclinic structure to the saddle-focus $O_{\mu}$. 
 
\begin{figure*}[!h]
\begin{center}
\includegraphics[height=7cm]{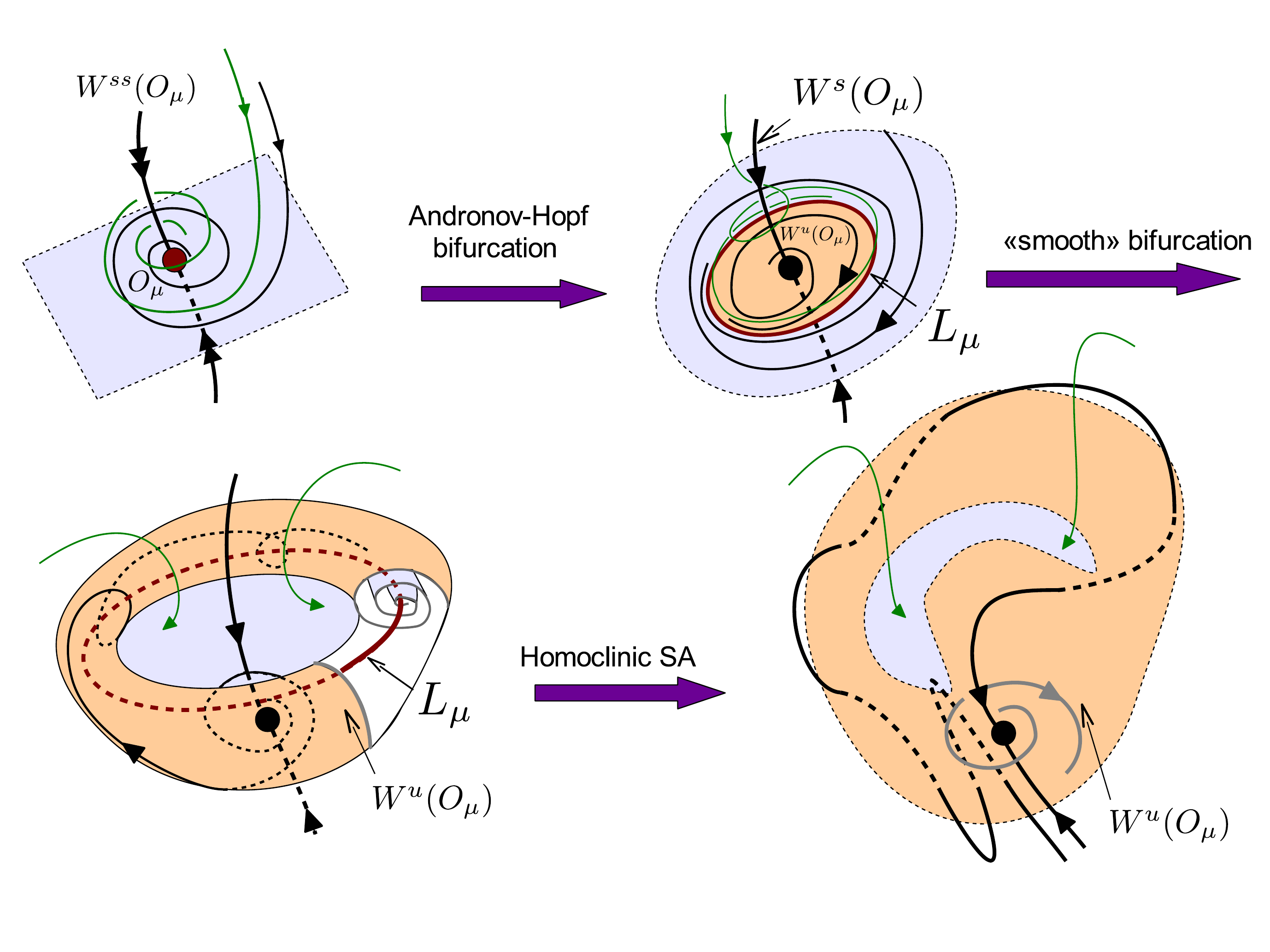}
\end{center}
\caption{Scenario of a discrete Shilnikov attractor creation. Reproduced from \cite{GonGon2016} with a permission of S.~V. Gonchenko.}
\end{figure*}

\section{Numerical evidences for a discrete Shilnikov attractor}
\label{num_bif}

In \cite{AshTownOrosz2007} the following system of $N$ phase oscillators was considered: 
$$ \dot{\theta}_n = \omega_n + \frac{1}{N}\sum\limits_{m=1}^{N} g(\theta_n - \theta_m) + \epsilon I_n(t) + \eta w_n(t),\, n=1,\dots, N,$$
where  $\theta_n$ is the phase of $n$th oscillator, $\omega_n$ is the natural frequency of the $n$th oscillator, $I_n (t)$ is an impulsive input with unit magnitude, $w_n(t)$ is uncorrelated white noise such that the associated random walk has unit growth of variance per unit time, and $g(\theta)$ is a coupling function.
If we suppose that all oscillators are identical (i.e., $\omega_n = \omega$ for all $n=1, \dots, N$) and there is no external inputs or any noise, then the system becomes simpler and takes form 
$$ \dot{\theta}_n = \omega + \frac{1}{N}\sum\limits_{m=1}^{N} g(\theta_n - \theta_m), \,n=1,\dots, N.$$
For any number of oscillators $N$ or coupling function $g(\theta)$ the dimension of this system could be reduced \cite{AshSwift1992} by introducing the phase differences $\gamma_n = \omega_n - \omega_{n^\ast}$, where $n^\ast \in \{ 1, \dots, N \}$ and could be any of these numbers. 
Following \cite{AshTownOrosz2007}, let us choose $N = 5$, $n^\ast = 5$, and $g(\theta) = -\sin{(\theta + \alpha)} + r \sin{(2\theta + \beta)}$.  
This gives us a system of five identical globally coupled phase oscillators with the biharmonic coupling. From \cite{AshTownOrosz2007} it is known that for some parameter values the plane $\gamma_3 = \omega_3 -\omega_5 = 1.2$ is a Poincar{\'e} section for the ODEs that describe this system.

Here we describe the sequence of Poincar{\'e} map's bifurcations that implies an existence of a discrete Shilnikov attractor in this system. 
Let us fix parameters $\alpha = 1.6,\, \beta = -1.58$ and vary parameter $r$. 
For $r < 0.1795$ a stable fixed point with two complex-conjugated multipliers is an attractor. At $r \approx 0.1795$ Neimark-Sacker bifurcation occurs: the fixed point becomes a saddle-focus and an attracting invariant curve appears. 
When we increase the parameter $r$ further, an unstable manifold of the saddle-focus fixed point starts to wind around the invariant curve. After that we observe a sequence of torus doublings; figures \ref{num_bif}-b) and \ref{num_bif}-c) show the first and the second doublings respectively. At some parameter value we see that the largest Lyapunov exponent becomes positive and we observe a {strange attractor}. 
If we continue increasing parameter $r$ we observe that the chaotic attractor becomes closer and closer to the saddle-focus (visually, at least). A graph of distance from the attractor to the saddle-focus supports this claim and shows that at $r = r_{\rm min}$ the distance is minimal. We computed invariant manifolds of saddle-focus, analyzed their relative positions at values of $r$ close to $r_{\rm min}$ and numerically found an intersection of invariant manifolds of the saddle-focus. Thus, we observe a sequence of bifurcations that satisfies the scenario of a discrete Shilnikov attractor emergence.  

\begin{figure}[!ht]
\begin{minipage}[h]{0.32\linewidth}
\center{\includegraphics[width=1\linewidth]{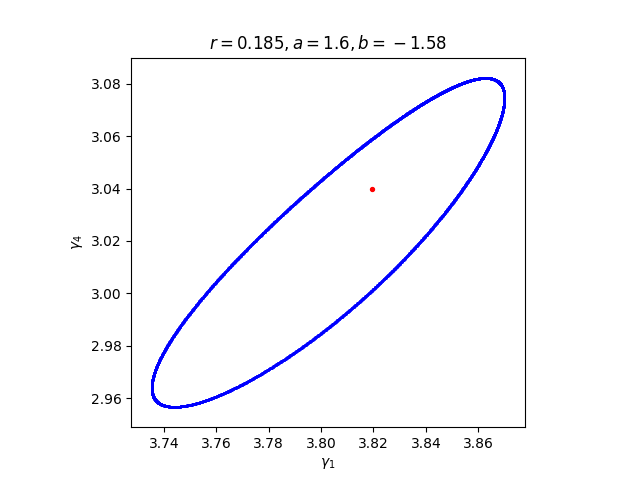} \\ a) {$r = 0.185$}}
\end{minipage}
\hfill
\begin{minipage}[h]{0.32\linewidth}
\center{\includegraphics[width=1\linewidth]{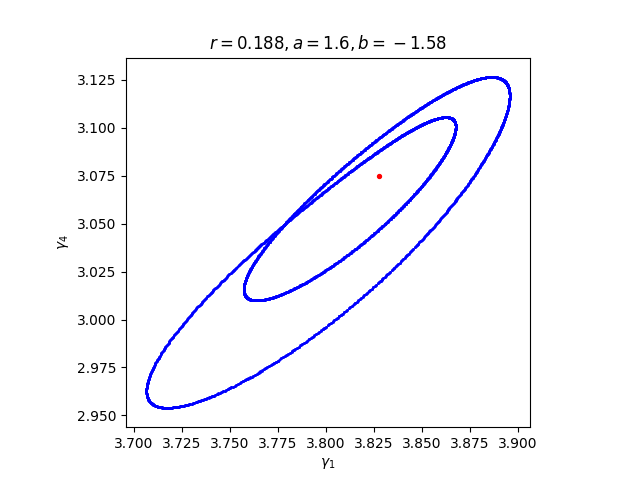} \\ b) {$r = 0.188$}}
\end{minipage}
\hfill
\begin{minipage}[h]{0.32\linewidth}
\center{\includegraphics[width=1\linewidth]{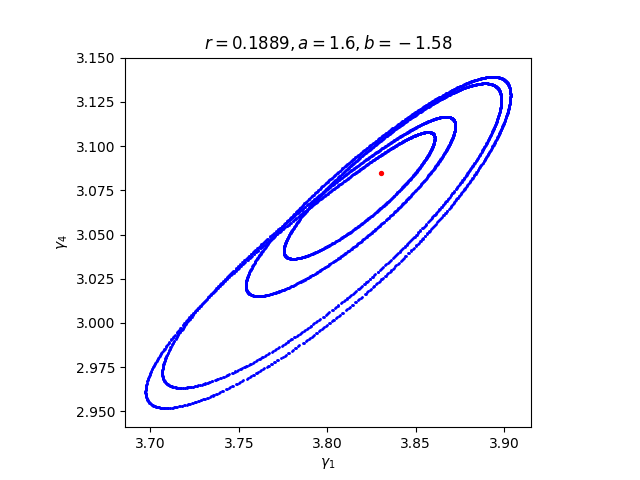} \\ c) {$r = 0.1889$}}
\end{minipage}
\vfill 
\begin{minipage}[h]{0.32\linewidth}
\center{\includegraphics[width=1\linewidth]{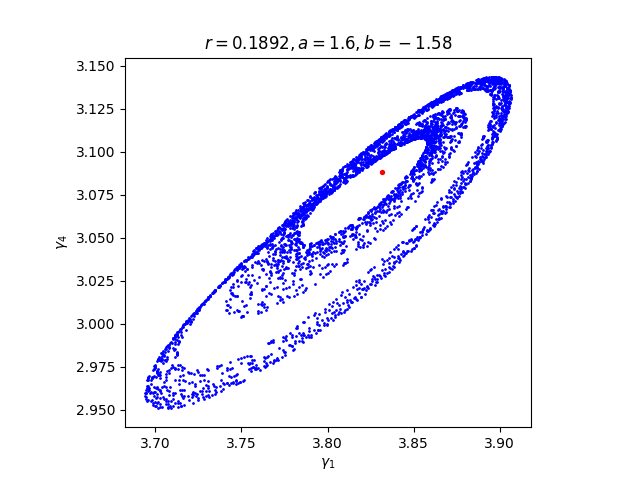} \\ d) {$r = 0.1892$}}
\end{minipage}
\hfill
\begin{minipage}[h]{0.32\linewidth}
\center{\includegraphics[width=1\linewidth]{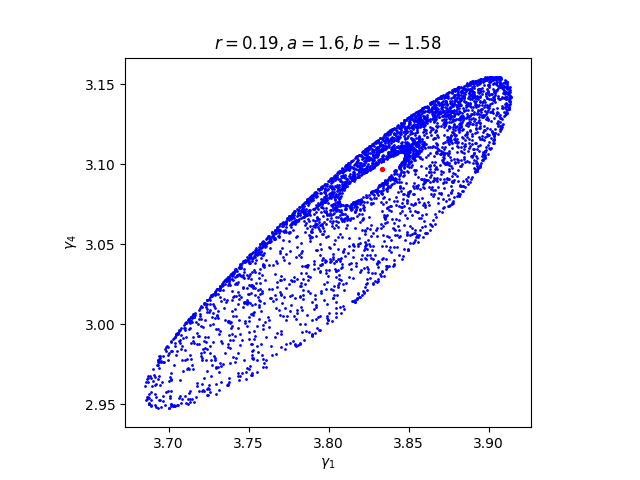} \\ e) {$r = 0.19$}}
\end{minipage}
\hfill
\begin{minipage}[h]{0.32\linewidth}
\center{\includegraphics[width=1\linewidth]{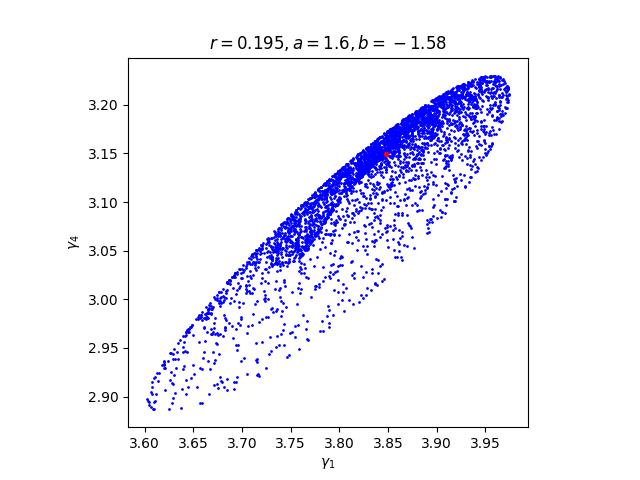} \\ f) {$r = 0.195$}}
\end{minipage}
\vfill 
\begin{minipage}[h]{0.32\linewidth}
\center{\includegraphics[width=1\linewidth]{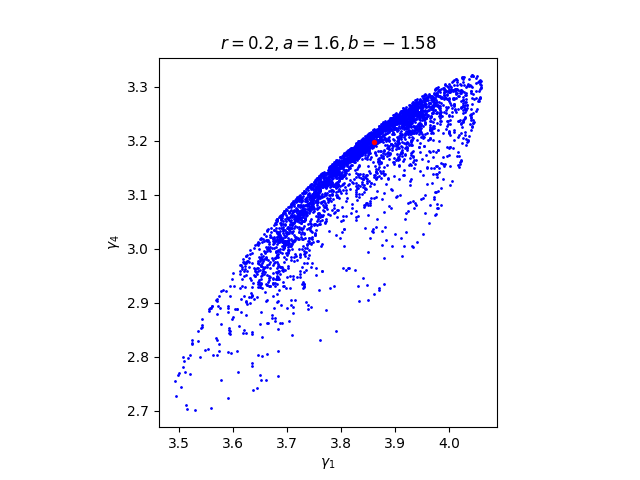} \\ g) {$r = 0.2$}}
\end{minipage}
\caption{Projections of the attractor (blue) and the saddle-focus fixed point (red) on the plane $(\gamma_1, \gamma_4)$ for the particular values of parameter $r$: a) -- the invariant curve; b) -- first torus doubling; c) -- second torus doubling; d) -- a strange attractor after a torus breakdown; e, f) -- increasing $r$ decreases the distance between the attractor and the fixed point; g) -- an attractor which looks like a symmetric copy of the chaotic attractor from \cite{AshTownOrosz2007}}
\end{figure}

\begin{figure*}[!ht]
\begin{center}
\includegraphics[height=5cm]{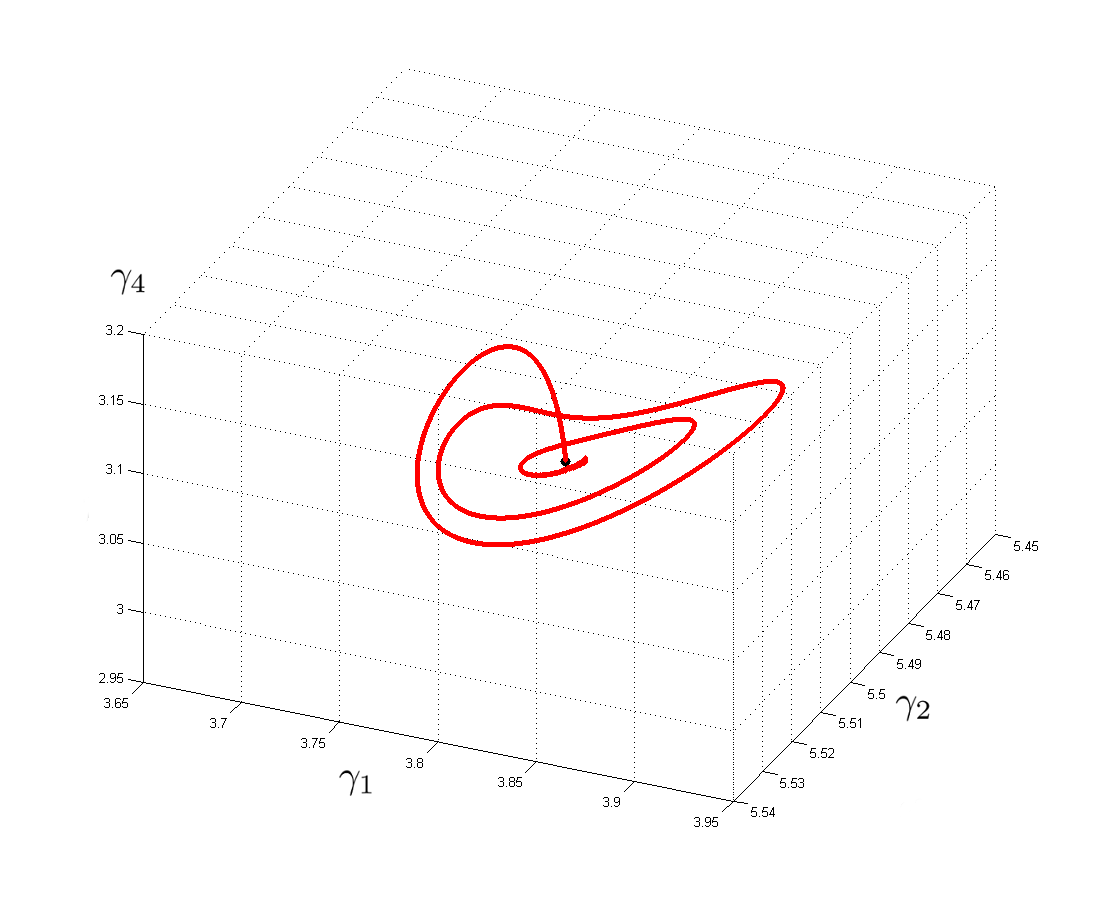}
\end{center}
\caption{A homoclinic structure to the saddle-focus fixed point.\label{hom_structure}}
\end{figure*}

The homoclinic structure of this discrete Shilnikov attractor has a one peculiar feature. Fig.~\ref{hom_structure} shows how one-dimensional unstable manifold of the saddle-focus looks like when it intersects with the stable manifold. Note that while this is a homoclinic structure to the saddle-focus of a three-dimensional map it looks more like a homoclinic trajectory to a saddle-focus of a three-dimensional flow. It is quite surprising because if a fixed point of a generic map has a homoclinic trajectory then its invariant manifolds have strong transverse oscillations in the neighbourhood of this fixed point. In present work we don't explore why exactly this happens here. Note that similar situation occured in \cite{BorKazSat2016}. The explanation for this behaviour was that some iteration of a map in the neighbourhood of a fixed point can be embedded into the flow of the same dimension hence the unconventional homoclinic structure of a fixed point.

\begin{figure*}[!ht]
\begin{center}
\includegraphics[height=5cm]{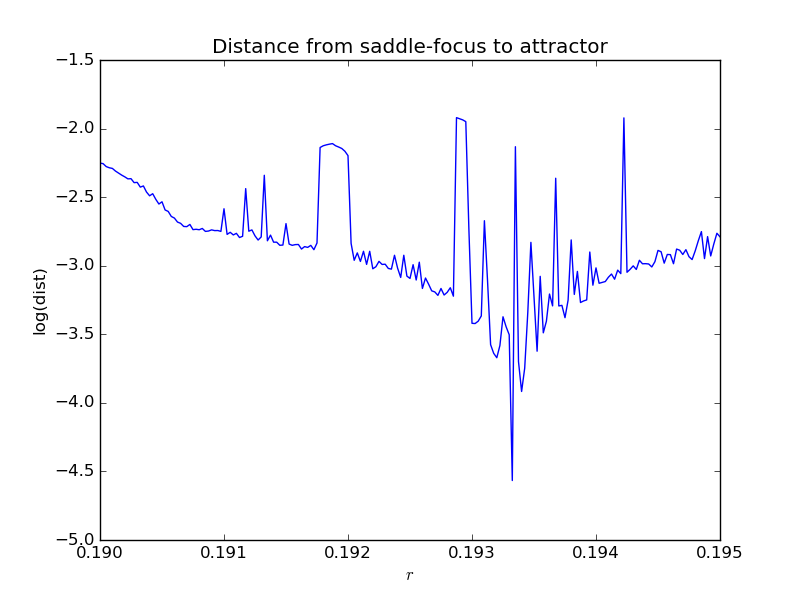}
\end{center}
\caption{Graph of distance from the saddle-focus to the attractor\label{dist_graph}}
\end{figure*}

\section{Discussion}

In present work we studied strange attractors that were found in the system of five identical globally coupled phase oscillators with the biharmonical coupling. We showed that homoclinic attractors based on a saddle-focus fixed point exist in this model. We also showed that these strange attractors appear due to scenarios that were suggested in \cite{GonGonShil2012,GonGonKazTur2014}. 

However, there are still some unresolved questions regarding the dynamics of this system. Why the homoclinic structure to the saddle-focus fixed point has this flow-like behaviour? Is it possible for this system to have homoclinic attractors based on a saddle fixed point with real multipliers? These questions would be the basis for our future research on this topic.

\section{Acknowledgements}
This paper was supported by Russian Science Foundation (Contract no. 14-12-00811). The numerical analysis that was presented in section \ref{num_bif} had been carried up with a support of Russian Science Foundation (Contract no. 17-11-01041). 

\bibliographystyle{alpha}
\bibliography{refs}

\newcommand{\etalchar}[1]{$^{#1}$}
\begin{thebibliography}{AOWT07}

\bibitem[ACS83]{ArnColSp1983}
A.~{Arneodo}, P.~H. {Coullet}, and E.~A. {Spiegel}.
\newblock Cascade of period doublings of tori.
\newblock {\em Physics Letters A}, 94:1--6, February 1983.

\bibitem[ACT80]{ArnCoulTres1980}
A~Arneodo, P~Coullet, and C~Tresser.
\newblock {Occurence of strange attractors in three-dimensional Volterra
  equations}.
\newblock {\em Physics Letters A}, 79(4):259--263, 1980.

\bibitem[ACT82]{ArnCoulTres1982}
A.~Arneodo, P.~Coullet, and C.~Tresser.
\newblock {Oscillators with chaotic behavior: An illustration of a theorem by
  Shil'nikov}.
\newblock {\em Journal of Statistical Physics}, 27(1):171--182, 1982.

\bibitem[Ani86]{Anish}
V.~S. Anishchenko.
\newblock Stochastic oscillations in radiophysical systems.
\newblock {\em Saratov: Saratov Gos. Univ.}, page~2, 1986.

\bibitem[AOWT07]{AshTownOrosz2007}
Peter Ashwin, Gábor Orosz, John Wordsworth, and Stuart Townley.
\newblock {Dynamics on Networks of Cluster States for Globally Coupled Phase
  Oscillators}.
\newblock {\em SIAM Journal on Applied Dynamical Systems}, 6(4):728--758, 2007.

\bibitem[AS92]{AshSwift1992}
P.~Ashwin and J.~W. Swift.
\newblock The dynamics of $n$ weakly coupled identical oscillators.
\newblock {\em Journal of Nonlinear Science}, 2(1):69--108, 1992.

\bibitem[BAR16]{AshBickRodr2016}
Christian Bick, Peter Ashwin, and Ana Rodrigues.
\newblock Chaos in generically coupled phase oscillator networks with
  nonpairwise interactions.
\newblock {\em Chaos: An Interdisciplinary Journal of Nonlinear Science},
  26(9):094814, 2016.

\bibitem[BKK17]{BakhKazKor2017}
Yu.~V. Bakhanova, A.~O. Kazakov, and A.G. Korotkov.
\newblock {Spiral chaos in Lotka-Volterra like models}.
\newblock {\em Zhurnal Srednevolzhskogo matematicheskogo obshchestva},
  19(2):13--24, 2017.

\bibitem[BKS16]{BorKazSat2016}
Alexey~V. Borisov, Alexey~O. Kazakov, and Igor~R. Sataev.
\newblock Spiral chaos in the nonholonomic model of a chaplygin top.
\newblock {\em Regular and Chaotic Dynamics}, 21(7):939--954, Dec 2016.

\bibitem[BTP{\etalchar{+}}11]{AshBickTimm2011}
Christian Bick, Marc Timme, Danilo Paulikat, Dirk Rathlev, and Peter Ashwin.
\newblock {Chaos in Symmetric Phase Oscillator Networks}.
\newblock {\em Phys. Rev. Lett.}, 107:244101, Dec 2011.

\bibitem[EM14]{EngelMir2014}
Jan~R. Engelbrecht and Renato Mirollo.
\newblock {Classification of attractors for systems of identical coupled
  Kuramoto oscillators}.
\newblock {\em Chaos: An Interdisciplinary Journal of Nonlinear Science},
  24(1):013114, mar 2014.

\bibitem[GG16]{GonGon2016}
AS~Gonchenko and SV~Gonchenko.
\newblock {Variety of strange pseudohyperbolic attractors in three-dimensional
  generalized H{\'e}non maps}.
\newblock {\em Physica D: Nonlinear Phenomena}, 337:43--57, 2016.

\bibitem[GGKT14]{GonGonKazTur2014}
Alexander Gonchenko, Sergey Gonchenko, Alexey Kazakov, and Dmitry Turaev.
\newblock Simple scenarios of onset of chaos in three-dimensional maps.
\newblock {\em International Journal of Bifurcation and Chaos}, 24(08):1440005,
  2014.

\bibitem[GGS12]{GonGonShil2012}
A.~S. Gonchenko, S.~V. Gonchenko, and L.~P. Shilnikov.
\newblock Towards scenarios of chaos appearance in three-dimensional maps.
\newblock {\em Rus. Nonlin. Dyn.}, 8:3--28, 2012.

\bibitem[HMM93]{HansMatoMeunier1993}
D.~Hansel, G.~Mato, and C.~Meunier.
\newblock Clustering and slow switching in globally coupled phase oscillators.
\newblock {\em Phys. Rev. E}, 48:3470--3477, Nov 1993.

\bibitem[KFR01]{KuznFeoRin2001}
Yu.~A. Kuznetsov, O.~De Feo, and S.~Rinaldi.
\newblock {Belyakov Homoclinic Bifurcations in a Tritrophic Food Chain Model}.
\newblock {\em {SIAM} Journal on Applied Mathematics}, 62(2):462--487, 2001.

\bibitem[KK01]{kori2001slow}
Hiroshi Kori and Yoshiki Kuramoto.
\newblock Slow switching in globally coupled oscillators: robustness and
  occurrence through delayed coupling.
\newblock {\em Physical Review E}, 63(4):046214, 2001.

\bibitem[New79]{Newhouse79}
Sheldon~E Newhouse.
\newblock The abundance of wild hyperbolic sets and non-smooth stable sets for
  diffeomorphisms.
\newblock {\em Publications Math{\'e}matiques de l'Institut des Hautes
  {\'E}tudes Scientifiques}, 50(1):101--151, 1979.

\bibitem[Oku93]{Okuda1993}
Koji Okuda.
\newblock Variety and generality of clustering in globally coupled oscillators.
\newblock {\em Physica D: Nonlinear Phenomena}, 63(3):424 -- 436, 1993.

\bibitem[PMT05]{PopMaistrTass2005}
Oleksandr~V. Popovych, Yuri~L. Maistrenko, and Peter~A. Tass.
\newblock Phase chaos in coupled oscillators.
\newblock {\em Phys. Rev. E}, 71:065201, Jun 2005.

\bibitem[PR15]{PikRos2015}
Arkady Pikovsky and Michael Rosenblum.
\newblock {Dynamics of globally coupled oscillators: Progress and
  perspectives}.
\newblock {\em Chaos: An Interdisciplinary Journal of Nonlinear Science},
  25(9):097616, 2015.

\bibitem[RM63]{RosMcArth1963}
M.~L. Rosenzweig and R.~H. MacArthur.
\newblock {Graphical Representation and Stability Conditions of Predator-Prey
  Interactions}.
\newblock {\em The American Naturalist}, 97(895):209--223, 1963.

\bibitem[Shi67]{Shil67}
L.~P. Shilnikov.
\newblock {On a Poincare–Birkhoff problem}.
\newblock {\em Math. USSR Sbornik}, 3:91--102, 1967.

\bibitem[Shi86]{Shilnikov86}
L.~P. Shilnikov.
\newblock Bifurcation theory and turbulence.
\newblock {\em Methods of the Qualitative Theory of Differential Equations,
  Gorky}, pages 150--163, 1986.

\bibitem[Sma65]{Smale65}
S.~Smale.
\newblock Diffeomorphisms with many periodic points.
\newblock {\em Differential and Combinatorial Topology}, 1965.

\bibitem[SPMS17]{StankPerMcClintStefa2017}
Tomislav Stankovski, Tiago Pereira, Peter V.~E. McClintock, and Aneta
  Stefanovska.
\newblock {Coupling functions: Universal insights into dynamical interaction
  mechanisms}.
\newblock {\em Rev. Mod. Phys.}, 89:045001, Nov 2017.

\bibitem[Win67]{Winfree1967}
Arthur~T. Winfree.
\newblock Biological rhythms and the behavior of populations of coupled
  oscillators.
\newblock {\em Journal of Theoretical Biology}, 16(1):15 -- 42, 1967.

\end{thebibliography}

\end{document}